\documentclass[conference]{IEEEtran}
\usepackage{cite}
\IEEEoverridecommandlockouts
\usepackage{amsmath,amssymb,amsfonts}
\usepackage{graphicx}

\usepackage{textcomp}
\usepackage{xcolor}
\def\BibTeX{{\rm B\kern-.05em{\sc i\kern-.025em b}\kern-.08em
    T\kern-.1667em\lower.7ex\hbox{E}\kern-.125emX}}

\usepackage{pifont}
\usepackage{url} 

\usepackage{amsmath,amsfonts}
\usepackage{array}
\usepackage[caption=false,font=normalsize,labelfont=sf,textfont=sf]{subfig}
\usepackage[hidelinks]{hyperref}
\usepackage{textcomp}
\usepackage{stfloats}
\usepackage{url}
\usepackage{flushend}
\usepackage{verbatim}
\usepackage{graphicx}
\usepackage{diagbox}
\usepackage{textcomp}
\usepackage{booktabs}
\usepackage{multirow}
\usepackage{graphicx}
\usepackage[normalem]{ulem}
\useunder{\uline}{\ul}{}
\usepackage{microtype}
\usepackage{paralist}
\usepackage{lscape}
\usepackage[table]{xcolor}
\usepackage{threeparttable} 

\usepackage{rotating}

\usepackage{array} 
\usepackage{booktabs} 
\usepackage{microtype}
\usepackage{algpseudocode}

\newcommand{\gpt}{\textsc{GPT-4}\@\xspace}
\newcommand{\llama}{\textsc{Llama3}\@\xspace}
\newcommand{\hermes}{\textsc{Hermes 2}\@\xspace}

\usepackage{xspace}
\newcommand{\tool}{\textsc{LLMorph}\@\xspace}
\newcommand{\numtests}{561,267\@\xspace}

\newcommand{\nummrtotal}{191\@\xspace}
\newcommand{\numpapers}{44\@\xspace}

\usepackage{eso-pic}
\usepackage{hyperref}

\hypersetup{
pdftitle={LLMORPH: Automated Metamorphic Testing of Large Language Models},
pdfauthor={Steven Cho; Stefano Ruberto; Valerio Terragni},
pdfsubject={This paper presents LLMORPH, an automated testing tool that leverages Metamorphic Testing to uncover faulty behaviors in Large Language Models without relying on human-labeled data.},
pdfkeywords={large language models; metamorphic testing;
machine learning testing; NLP; Software Engineering for AI}
}

\begin{document}

\AddToShipoutPictureFG*{%
  \AtPageLowerLeft{%
    \raisebox{1.6cm}{%
      \hspace*{3.0cm}%
      \parbox{\dimexpr\paperwidth-5.0cm\relax}{%
        \centering\small
        This is the authors' accepted manuscript of the paper published in 
        \textit{40th IEEE/ACM International Conference on Automated Software Engineering (ASE 2025)}.
        The version of record is available via DOI: \href{https://doi.org/10.1109/ASE63991.2025.00385}{10.1109/ASE63991.2025.00385}.%
      }%
    }%
  }%
}
\title{\tool: Automated Metamorphic Testing \\ of Large Language Models}
\author{
    \IEEEauthorblockN{Steven Cho}
    \IEEEauthorblockA{
        University of Auckland\\
        Auckland, New Zealand\\
        steven.cho@auckland.ac.nz
    }
    \and
    \IEEEauthorblockN{Stefano Ruberto}
    \IEEEauthorblockA{
        JRC European Commission\\
        Ispra, Italy\\
        stefano.ruberto@ec.europa.eu
    }
    \and
    \IEEEauthorblockN{Valerio Terragni}
    \IEEEauthorblockA{
        University of Auckland\\
        Auckland, New Zealand\\
        v.terragni@auckland.ac.nz
    }
    }

\maketitle

\begin{abstract}

Automated testing is essential for evaluating and improving the reliability of Large Language Models (LLMs), yet the lack of automated oracles for verifying output correctness remains a key challenge. We present \tool, an automated testing tool specifically designed for LLMs performing NLP tasks, which leverages Metamorphic Testing (MT) to uncover faulty behaviors without relying on human-labeled data. MT uses Metamorphic Relations (MRs) to generate follow-up inputs from source test input, enabling detection of inconsistencies in model outputs without the need of expensive labelled data. \tool is aimed at researchers and developers who want to evaluate the robustness of LLM-based NLP systems. In this paper, we detail the design, implementation, and practical usage of \tool, demonstrating how it can be easily extended to any LLM, NLP task, and set of MRs. In our evaluation, we applied 36 MRs across four NLP benchmarks, testing three state-of-the-art LLMs: GPT-4, LLAMA3, and HERMES 2. This produced over 561,000 test executions. The results demonstrate \tool's effectiveness in automatically exposing incorrect model behaviors at scale. The tool source code is available at \url{https://github.com/steven-b-cho/llmorph}. A screencast demo is available at \url{https://youtu.be/sHmqdieCfw4}.




\end{abstract}

\smallskip

\section{Introduction}


\textbf{Large Language Models (LLMs)} have seen widespread adoption across various domains thanks to their strong performance in natural language understanding and generation tasks. Despite this success, serious concerns remain regarding their reliability and trustworthiness (e.g., bias, hallucination)~\cite{chang2024survey}. Identifying these problems is a critical prerequisite to effectively addressing and mitigating them. Therefore, automated testing of LLM is crucial for evaluating the quality of LLM outputs.
Moreover, automated testing of LLMs is becoming more important in industry, as many companies are moving away from using public LLM services and instead are running their own fine-tuned models locally~\cite{hanke2024open}. This change is often due to concerns about privacy, security, and legal rules, or simply the need for better results in their specific tasks~\cite{hanke2024open}.

\smallskip
One of the primary challenges in automatically testing LLMs is the \textbf{oracle problem}~\cite{barr2014oracle}; \textit{the problem of distinguishing between correct and incorrect test executions}. In \textbf{Natural Language Processing (NLP)}, generating new test inputs is relatively easy because large amounts of text data are readily available. However, checking whether the model’s outputs are correct given an  NLP task is much harder. This task usually relies on human-annotated labels, which act as ``oracle'' to judge correctness. Since manually creating these labeled datasets is time-consuming and costly, there is a strong need for automated oracles that can evaluate output quality without depending on human-generated labels.

\smallskip
\textbf{Metamorphic Testing (MT)}\cite{2017-chen-cs} is a widely used technique for addressing the oracle problem in software testing. Instead of requiring expected results for every test case, MT relies on \textbf{Metamorphic Relations (MRs)}: expected relationships among the outputs of related inputs. The key idea is that, even if one cannot automatically determine whether a single output is correct, \textit{we can use the relationships among the expected outputs of multiple related inputs as a test oracle}~\cite{2017-chen-cs}. An MR describes how the output should change (or remain consistent) when the input is modified in a specific way. For example, if two input texts are paraphrases of one another, then their outputs should also be similar. If this expected relationship is violated, the system may have produced an incorrect result.

\smallskip
To apply MT in practice, testing begins with a source input. This input is then transformed to create one or more follow-up inputs that satisfy the conditions of a given MR. The system under test is executed on both the original and the transformed inputs, and their outputs are compared. If the outputs do not satisfy the expected relationship defined by the MR, the test reports a failure. Thus, MT can detect faults even though the exact ground truth labels are not required for either case. A key advantage of MT is that the same MRs can be applied to many different inputs, enabling automated testing at large scale. This is particularly useful for testing LLMs, where failures often emerge only under specific input conditions~\cite{chang2024survey}, and where large volumes of unlabeled data are available.

\smallskip
Although MT has been applied across many NLP
tasks,
its use with LLMs remains relatively understudied~\cite{mr-metal}. To fill this gap, we recently presented the most comprehensive study on MT for LLMs in NLP to date. Our study was recently accepted at ICSME 2025~\cite{cho2025metamorphic}. We conducted the first systematic literature search on MRs for NLP, reviewing 1,024 papers and identifying \numpapers that explicitly defined MRs. This resulted in a catalog of \nummrtotal unique MRs across 24 tasks. We also presented \textbf{\tool}, an automated test tool for MT on LLMs, implementing 36 of the 191 MRs across four tasks.

\smallskip
\tool was originally produced as a means to conduct our study~\cite{cho2025metamorphic}. In this demonstration paper, we focus specifically on the tool itself, giving details on its design, implementation and usage. The intended users of \tool are researchers and developers who wish to verify the robustness of an LLM system for the purpose of verification or improvement. Given an LLM and a list of test inputs, \tool produces a list of failing metamorphic test pairs, allowing users to identify potentially unknown faults. The source code of \tool can be found at: \texttt{\url{https://github.com/steven-b-cho/llmorph}}



\begin{figure*}
    \centering
    \includegraphics[width=0.9\linewidth]{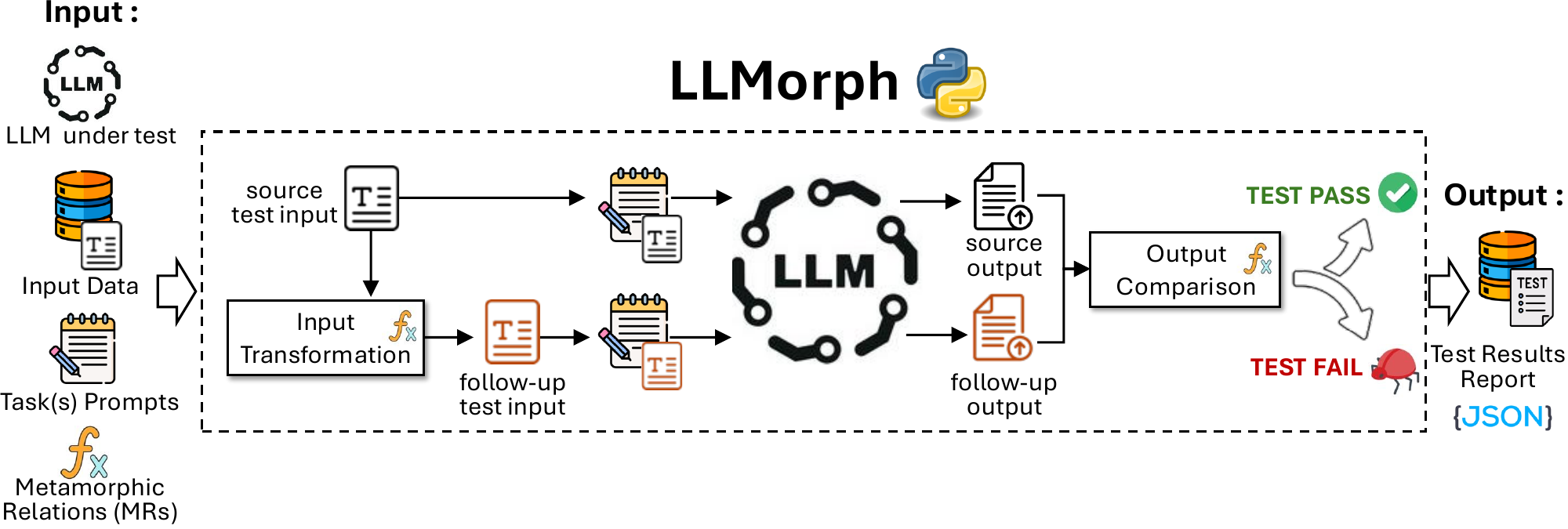}
    \vspace{-5mm}
    \caption{Logical architecture of \tool}
    \label{fig:tool}
\end{figure*}

\section{LLMorph}






The tool leverages metamorphic testing to measure the robustness of an LLM system. It enables users to apply metamorphic relations to any natural language task of their choice. Figure~\ref{fig:tool} illustrates the logical architecture of the tool. 

\subsection{Input}

\tool takes as input four items:

\smallskip
\noindent
\textbf{LLM under test.} Currently, \tool supports LLMs via the \textsc{OpenAI API}, and thus it is not necessary to host a local model. However, the tool can easily be extended to work with other APIs as well as locally deployed LLMs, if desired.

\smallskip
\noindent
\textbf{Input data.}
The input test data is structured as a list of textual source inputs of arbitrary length. Unlike in traditional testing, the inputs in \tool do not need to be labelled, as it is a metamorphic testing tool. This allows for a much larger amount of data to be used, since the cost associated with labelling the data is no longer a concern.

\smallskip
\noindent
\textbf{Task(s) prompts.} The user can choose which NLP tasks to test the LLM on. \tool currently supports four built-in tasks: context-based question answering (QA), natural language inference (NLI), continuous sentiment analysis (SA), and relation extraction (RE). These tasks are implemented using zero-shot prompts to the LLM under test. Users can easily extend the tool by adding new tasks or modifying the existing ones through simple JSON files, which define the prompts used for each task.
This is particularly useful when testing a fine-tuned LLM on any arbitrary task.

\smallskip
\noindent
\textbf{Metamorphic Relations (MRs).} The user can choose which MRs to test the LLM using. Some relations can be used in multiple tasks, while others are specific to only one. In our recent work, we performed a systematic literature review and identified 191 MRs for NLP~\cite{cho2025metamorphic}; of these, \tool currently implements 36 (see~\cite{cho2025metamorphic} for more details). This subset was chosen based on commonality, applicability across different tasks, specificity to individual tasks, and ease of understanding.


\subsection{Output} \tool produces as output a test result report, which is a JSON file containing information about all test runs. It includes the original source inputs, the generated follow-up inputs, the corresponding source and follow-up outputs produced by the LLM under test, and whether the metamorphic relation was violated or not in each case.

\subsection{Process} 

For each combination of source input, task, and MR, \tool uses the input transformation specified by the MR to produce a follow-up input. Both inputs are processed through the same LLM under test using the task prompt to produce the respective source and follow-up outputs. These are then compared to determine whether they satisfy the output relation specified by the MR.

\smallskip
\noindent
\textbf{Example.}
Given \textsc{GPT-4o} as the \textbf{LLM under test}, and the \textbf{source test input}:
\texttt{"The area in which a glacier forms is called a cirque. What geological features formed by glaciers?"},
let us assume the user selects the Question Answering (QA) \textbf{task} and the \textbf{metamorphic relation}:
\texttt{"Adding random spaces to the input should not change the output."}

\smallskip
\tool begins by applying the transformation associated with the MR, resulting in the follow-up input:
\texttt{"Th e a rea in wh ich a gl acier forms is cal l ed a ci rque. Wha t geologi cal f eatures form ed by glaci ers?"}

\smallskip
The QA task is defined by the following prompt:
\begin{quote}
\texttt{Here is some information: "{INPUT\_0}" Using only this information, nothing else, answer the following question: "{INPUT\_1}" Keep your answer to a short sentence. If you cannot give an answer, write 'unknown'.}
\end{quote}

Given this prompt, the LLM is queried twice: once with the original input and once with the transformed input. The \textbf{source output} is \texttt{unknown}, the \textbf{follow-up output} is \texttt{cirque}.

\smallskip
The \textbf{output comparison} based on the MR expects that both outputs should be the same. Since they differ, the MR is violated, and a faulty execution is detected.
Note that we can find this real fault in GPT-4o without the need for labeled data specifying that \texttt{unknown} should be the correct output\footnote{A cirque is a geological feature that \textit{forms} glaciers, not \textit{formed by} them. The correct answer should be unknown}.

\smallskip
This simple example illustrates how \tool automatically detects bugs in LLMs using MT. \tool supports a broader set of more complex (and potentially more insightful) MRs across three additional, more advanced NLP tasks. A key strength of \tool is its extensibility: additional metamorphic relations and tasks can be easily incorporated by simply adding to corresponding files.







\subsection{Implementation Details}

\tool is implemented as a \textsc{Python3} project. It uses the \texttt{openai} library for communication with the LLM, the \texttt{sentence\_transformers} library for semantic similarity calculation, and the \texttt{nlpaug} library to implement some MRs.

\smallskip
The implementation of the MRs are, on the whole, derived from semantic, non-precise definitions. The relations implemented currently have the following properties: An \textit{input transformation} function that takes a source input and transforms it to create a follow-up output; a \textit{output comparison} function that takes two outputs and compares them to determine whether they satisfy the MR's output relation; and (optionally) a set of \textit{verifications}, which are restrictions to the inputs and outputs that must hold for the MR to be valid. This latter property is not how MRs are typically definitionally structured; however, it was done this way for ease of implementation.
We implement the MRs through both function-based and LLM-based means. Simpler relations, such as \textit{MR-84: Concatenating a random sentence}, are done through traditional functions, or a library. More complex MRs -- for instance, \textit{MR-51: Paraphrasing} -- are implemented using a few-shot prompted LLM. The use of LLMs to test LLMs is a commonly used technique, enabling processes which would be much harder to achieve through traditional means. This transformation LLM is not (necessarily) the LLM under test, and is specified in the configuration file.

\smallskip
For tasks with multiple inputs (e.g., \textit{premise} and \textit{hypothesis} in NLI), if applicable, the relation will be applied to as many combination of inputs as possible. For instance, a relation could be applied to an NLI input's \textit{premise}; to its \textit{hypothesis}; or to both at once. This may result in multiple possible follow-up inputs from a single source input and MR.

\smallskip
Syntactic output comparisons may be done by direct equivalence, difference, set comparison, or another method, depending on the MR.
Semantic output comparisons are done using cosine similarity via the BERT-based model \textsc{paraphrase-MiniLM-L6-v2}. We specify similarity thresholds to determine (depending on the MR) if two outputs are equivalent (0.8 by default) and different (0.4 by default).
Numerical equivalence is determined by direct comparison with a 0.1 error window.


\section{Tool Usage}



\textbf{Installation.} \tool requires \textsc{Python 3.10}. Dependancies are installed via \texttt{python install -r requirements.txt}. An OpenAI API key should be put into \path{security/token-key.jwt}. This key is used for both the LLM under test and the LLM-based MRs.


\subsection{Running the tool}

\tool can be run using two methods: through a command line interface (CLI), or through a configuration file where we can set more parameters. 

\smallskip
\textbf{CLI.} To run using CLI, execute \texttt{python src/main.py} with the the following arguments:

\begin{itemize}
    \item \texttt{llm}: The ID of the LLM to test. This is the model name to be sent through the OpenAI API.
    \item \texttt{task}: The name of the NLP task to test on. The list of tasks can be found in \path{src/config/list_tasks.json}.
    \item \texttt{mr}: The name of the metamorphic relation to test using. The list of relations can be found in \path{src/config/list_relations.json}.
    \item \texttt{input\_data}: The path to the JSON file containing the inputs. Structured as an array of data points.
    \item \texttt{base\_dir}: The path to the directory where caches and outputs will be stored.
\end{itemize}

\smallskip
\textbf{Configuration file.} To run using the config file (found at \path{src/config/run_config.json}), execute \texttt{python src/mt\_main.py}. Some notable parameters include:

\begin{itemize}
    \item \texttt{llm\_list}: A list of LLM names. All LLMs in the list will be tested.
    \item \texttt{tasks}: A dictionary of lists, with task name as the key and a list of MRs as the value. An empty list will run all available MRs.
    \item \texttt{checkpoint\_interval}: The number of instances processed before saving a checkpoint.
    \item \texttt{continue\_from\_checkpoint}: If \texttt{true}, will continue run from the checkpoint with the latest creation time.
    \item \texttt{llm\_endpoint}: The endpoint for the LLM under test, as well as the LLM for MR implementation. Change if not using the default OpenAI endpoint.
    \item \texttt{llm\_for\_transformation}: The LLM used for MR implementation. Defaults to the LLM under test.
\end{itemize}

More parameter details can be found in \texttt{README.md}. 


\subsection{Reading the output}

The results can be found in \path{{base_dir}/results} (with \path{base_dir} being specified in the run parameters). The output is a JSON file with a list of responses, with each instance specifying the following:

\begin{itemize}
    \item \texttt{source\_input}: The original test input.
    \item \texttt{source\_output}: The response of the LLM from the source input.
    \item \texttt{followup\_inputs}: A list containing the transformed input(s).
    \item \texttt{followup\_outputs}: A list containing the responses of the LLM from the follow-up input(s).
    \item \texttt{relation}: A list specifying whether the follow-up output(s) satisfied the output relation.
    \item \texttt{verification\_failure}: Whether the metamorphic group was valid according to any restrictions specified for the metamorphic relation.
\end{itemize}

\subsection{Adding and modifying tasks, relations, and LLMs}

\textbf{Tasks.} The currently implemented NLP tasks are done via prompts to the LLM. These prompts can be modified, and new tasks added, through editing \path{src/config/list_tasks.json} and \path{src/config/template/sut_prompt_templates.json}.

\smallskip
\textbf{Metamorphic relations.} 
The currently implemented MRs are done via either traditional function implementation, or prompts to an LLM.
These can be modified, and new relations added, through editing the following: \path{src/relations/func_it.py} and \path{src/relations/func_or.py} for the implementation of the input transformation and output relation, respectively; \path{src/config/template/it_prompt_templates.json} or \path{src/config/template/or_prompt_templates.json} if using a prompted LLM for transformation or comparison; and \path{src/config/list_relations.json} to specify the particular MR.


\smallskip
\textbf{LLMs.} Currently, \tool supports LLMs using the OpenAI API. This is controlled by the \texttt{llm\_list} and \texttt{llm\_endpoint} parameters in the config file. To use another API, or to run a local model, modify \path{src/llm_runner.py}.

\section{Evaluation}

To evaluate the ability for \tool to detect faulty behaviour, we conducted large-scale experiments using \tool on three popular LLMs (\gpt, \llama, and \hermes) and four datasets (\textsc{SQuAD2}, \textsc{SNLI}, \textsc{SST2}, and \textsc{RE-DOCRED}), leading to \numtests test executions. We found that \tool effectively exposes faulty behaviours, with an average failure rate of 18\%. 
Comparing with traditional testing with hard-to-obtain labelled data, we found it complementary, with MT being able to detect bugs not found by the former. We also manually analysed 937 metamorphic oracle violations and found 
a range of false positive rates, varying from 0\% to 70\%, depending on the MR and task.
Most of these arise from the intrinsic limitations of MT for NLP and aligns with traditional MT for NLP~\cite{cho2025metamorphic}.
In addition, we found that the evaluation of each input was fast, usually depending only on the speed of the 2-3 calls to the LLM per source input.

\smallskip
\emph{For more information, please refer to our ICSME paper~\cite{cho2025metamorphic}, which describes the evaluation in detail.}

\section{Related Work}


Traditionally, testing of LLMs use benchmarks (e.g., MMLU \cite{hendryckstest2021}), where the output of the LLM is compared to ground truths to determine the LLM's effectiveness in the area of test. However, this requires labelled data, which is costly to obtain. \tool instead utilises Metamorphic Testing, which does not require any labelled data, allowing one to use fully automated testing using cheaper methods.




\smallskip
Hyun et al.~\cite{mr-metal} recently presented \textsc{METAL}, a framework for MT for LLMs on NLP tasks.
There are some limitations to their work, however, which we address in ours. They investigate 13 MRs, while we implement 36; they use zero-shot prompting in their LLM MR implementation, while we use few-shot; we have a CLI availability as well as a large number of configuration parameters, while they do not; our users can choose what MR to run, while theirs cannot; in ours it is easy to add new tasks and relations, while in theirs it is not; and ours is fully realised and modular, rather than a simple Jupyter Notebook.

\section{Conclusion and Future Work}



Despite the vast resources poured into producing and testing large language models, using Metamorphic Testing in this area is still unexpectedly underexplored. \tool is one of the first to explore this space, providing a foundation from which to investigate MT for LLMs.


\tool currently implements 36 out of the 191 Metamorphic Relations (MRs) we have collected~\cite{cho2025metamorphic}. By open-sourcing the tool and designing it to be modular and easily extensible, we hope the community will contribute to expanding its support for additional MRs and NLP tasks.

\bibliographystyle{IEEEtran}
\bibliography{bibliography}

\end{document}